\documentclass[12pt,tightenlines,eqsecnum,floats,showpacs,nofootinbib,amsmath,amssymb,aps,prd]{revtex4}

\usepackage{graphicx}
\usepackage{amsmath,amssymb}
\usepackage{subfigure}

\def\be{\begin{equation}}
\def\ee{\end{equation}}
\def\ba{\begin{eqnarray}}
\def\ea{\end{eqnarray}}

\def\f{\frac}
\def\tf{\tfrac}
\def\dd{{\rm d}}
\def\sgn{\mathrm{sgn}}
\def\d{{\rm d}}
\def\la{\lambda}
\def\g{{\rm grav}}
\def\e{\mathring{e}}
\def\w{\mathring{\omega}}
\def\q{\mathring{q}}
\def\xiz{\mathring{\xi}}
\def\epz{\mathring{\epsilon}}

\def\lp{\ell_{\mathrm{Pl}}}
\def\m{{\rm matt}}
\def\ve{\varepsilon}
\def\Th{\Theta}
\def\R{\mathbb{R}}
\def\Z{\mathbb{Z}}
\def\H{\mathcal{H}}
\def\Hkg{\mathcal{H}_{\rm kin}^{\rm grav}}

\begin{document}

\title{Loop quantum cosmology of Bianchi type IX models}

\author{Edward Wilson-Ewing}
\email{wilsonewing@gravity.psu.edu}
\affiliation{Institute for Gravitation and the Cosmos,
Physics Department,\\ The Pennsylvania State University,
University Park, PA 16802, USA}

\begin{abstract}

The loop quantum cosmology ``improved dynamics'' of the Bianchi type
IX model are studied.  The action of the Hamiltonian constraint
operator is obtained via techniques developed for the Bianchi type I
and type II models, no new input is required.  It is shown that the
big bang and big crunch singularities are resolved by quantum gravity
effects.  We also present the effective
equations which provide modifications to the classical equations of
motion due to quantum geometry effects.

\end{abstract}


\maketitle

\section{Introduction}
\label{s1}

Loop quantum cosmology (LQC) \cite{mbrev, aa-rev} is an approach to
quantum cosmology following the ideas of loop quantum gravity (LQG)
\cite{alrev, crbook, ttbook}.  One of the major results of LQC in
the homogeneous and isotropic Friedmann-Robertson-Walker (FRW)
models is that, while general relativity approximates the dynamics
very well in the low (with respect to the Planck scale)
curvature regime, the classical big bang singularity is avoided:
when the matter energy density approaches the Planck energy density,\
deviations from general relativity become significant and a ``quantum
bounce'' due to quantum gravity effects occurs when the matter energy
density reaches a critical energy density of the order of the the Planck
density \cite{mb1, abl, aps3, acs, apsv, warsaw1, rv, kv, ls}.  More
recently, it has been shown that the singularity is also resolved in
the improved dynamics approach of loop quantum cosmology in the
anisotropic Bianchi type I and type II cosmological models
\cite{awe2, awe3} and in the hybrid loop-Fock quantization of the
inhomogeneous Gowdy model \cite{gowdy}.  The goal of this paper is
to extend the LQC improved dynamics analysis of the Bianchi type I
and type II models to the more complicated Bianchi type IX models.

At the classical level, the Bianchi IX model has a much richer
phenomenology than Bianchi I and II models as it displays Mixmaster
dynamics as the singularity is approached \cite{bb}.  In essence, a
space-time which exhibits Mixmaster dynamics is one which can be
described for long periods of time (known as epochs) as a Bianchi
I space-time characterized by three anisotropic expansion rates.
Such a space-time will occasionally undergo a ``Mixmaster bounce''
from one epoch to another where the three expansion rates change
in a specific manner.  Bianchi I models approach the singularity
in a rather straightforward way as they do not undergo any Mixmaster
bounces while Bianchi II models may undergo a single Mixmaster bounce
between two epochs as the singularity is approached (see \cite{bb}
and references therein).  The Bianchi IX model, on the other hand,
undergoes many Mixmaster bounces and this behaviour is chaotic
\cite{bb, dhn}.  Since much of this behaviour occurs when the
curvature is of the Planck scale, quantum gravity effects cannot
be neglected and the Mixmaster behaviour may be significantly
modified when they are taken into account.

Bianchi IX models are also thought to play an important role near
generic singularities in classical general relativity.  The Belinskii,
Khalatnikov, Lifshitz (BKL) conjecture suggests that as a generic
space-like singularity is approached, time derivatives dominate over
spatial derivatives whence physical fields at each point evolve
independently from those at neighbouring points.  Dynamics can
therefore be approximated by the ODE's used in homogeneous space-times,
most generally a Bianchi IX solution with a massless scalar field
\cite{bkl1, bkl2}.  Since other matter fields do not contribute to
the dynamics as the singularity is approached, we will only consider
the case of a massless scalar field in this work.  There has recently
been a considerable amount of numerical work supporting this paradigm
(see, e.g., \cite{bb}) and the conjecture has also been rewritten in
terms of variables suitable to a loop quantization \cite{ahs}.  If the
BKL conjecture is indeed correct, it follows that a good understanding
of the quantum dynamics of the Bianchi IX model in the deep quantum
regime could lead to major insights into the behaviour of \emph{generic}
space-times in regions where the curvature reaches the Planck scale.

Because of the Bianchi IX model's importance, it has already been the
subject of studies within the framework of loop quantum cosmology,
both in a pre-$\mu_o$-type Hamiltonian framework \cite{bdv, bdh} and
in a spin-foam-inspired dipole cosmology model (first introduced for
the isotropic case in \cite{rv}) which also allows inhomogeneities
\cite{bmr}.  However, it is important to study the improved
$\bar\mu_i$-type dynamics of LQC since it has been shown that the
predictions of the pre-$\mu_o$ approach are unphysical in the infrared
limit.  In particular, in isotropic cosmological models quantum gravity
effects can become important at energy densities arbitrarily below the
Planck scale in this scheme.  To ensure that quantum gravity effects
only become important at the Planck scale, one must instead use the
improved dynamics approach in the Hamiltonian framework%
\footnote{It is possible that once the curvature reaches the Planck
scale a scheme other than $\bar\mu_i$ may be the correct one
but we will only consider the $\bar\mu_i$ scheme here.}.
On the other hand, since the dipole cosmology model presented in
\cite{bmr} is inspired by spin foam models, that approach is
complementary to ours and it will be interesting to compare the
results of these two frameworks.

As pointed out above, chaotic behaviour appears as the singularity is
approached in classical Bianchi IX space-times.  It has been argued
in the pre-$\mu_o$ LQC treatment of the model that this behaviour is
avoided in LQC due to quantum gravity effects \cite{bdh}.  In
essence, the argument is that quantum gravity effects become important
before a significant number of Mixmaster bounces occur.  Since the
quantum gravity effects are repulsive, the space-time will exit the
Planck regime having only undergone a small number of Mixmaster bounces
and hence the dynamics are not chaotic.  In this paper, we see that
in some cases this occurs already in the effective theory which
incorporates the quantum geometry effects into the dynamics.
However, we cannot yet show that this is a generic result.  We will
go into more detail in Sec.~\ref{s4}.

The outline of the paper is as follows.  In Sec.~\ref{s2}, we will
briefly review the necessary classical properties of the Bianchi type
IX model in order to proceed with the quantization.  In Sec.~\ref{s3}
we will study the quantum properties of the model, first recalling
the kinematics which are the same as for the Bianchi type I and type
II models studied in \cite{awe2, awe3}.  We will then study the
Hamiltonian constraint operator for the Bianchi IX model with a
massless scalar field as the matter field.  The Hamiltonian constraint
operator gives an evolution equation where the scalar field acts as
a relational time parameter.  The dynamics of the model are obtained
by using the same technology that was developed during the study
of the improved LQC dynamics of the Bianchi type I and type II models
\cite{awe2, awe3}; \emph{there is no need to introduce any new operators}.
In Sec.~\ref{s4} we will derive effective equations which
provide the first order quantum corrections to the classical equations
of motion and in Sec.~\ref{s5} we summarize our results and discuss
open issues.

\section{Classical Theory}
\label{s2}

In Bianchi models \cite{taub,bianchi,atu}, one restricts oneself to
those phase space variables which admit a 3-dimensional group of
symmetries which act simply and transitively.  The symmetries allowed
in the Bianchi IX group are the three spatial rotations on $S^3$.  It
follows that the three Killing (left invariant) vector fields $\xiz^a_i$
satisfy%
\footnote{Here we are following the convenctions used in \cite{bmr}.
A different, although equivalent, choice is used in \cite{apsv, bdv, bdh}
where the structure constants differ by an overall sign.}%
\be [\xiz_i, \, \xiz_j] = \f{2}{r_o} \epz^k{}_{ij} \xiz_k, \ee
where the structure constants are given by the completely antisymmetric
tensor $\epz_{ijk}$ times $2/r_o$ where $r_o$ is the radius of the
3-sphere with respect to the fiducial metric.  $\epz_{ijk}$ is defined
such that $\epz_{123} = 1$, note that the internal indices $i, j, k,
\ldots$ can always be freely raised and lowered.  There is also a
canonical triad $\e^a_i$ ---the right invariant vector fields--- which
is Lie dragged by $\xiz^a_i$. It is convenient to use $\e^a_i$ and its
dual co-triad $\w_a^i$ as fiducial frames and co-frames. They satisfy:
\be   [\e_i,\, \e_j] = - \f{2}{r_o} \epz^k{}_{ij}\, \e_k, \quad\quad \dd\,\w^k
= \f{1}{r_o}\, \epz^k{}_{ij} \w^i\wedge\w^j. \ee

The form of the equations above indicates that $M$ admits global
coordinates $\alpha \in [0, 2\pi), \beta \in [0, \pi)$ and $\gamma \in
[0, 4\pi)$ such that for $r_o = 2$ the Bianchi IX co-triads have the form
\ba \w_a^1 &=& \sin\beta\sin\gamma (\dd\alpha)_a + \cos\gamma
(\dd\beta)_a, \nonumber \\ \w_a^2 &=& -\sin\beta\cos\gamma
(\dd\alpha)_a + \sin\gamma (\dd\beta)_a, \\ \w_a^3 &=&
\cos\beta (\dd\alpha)_a + (\dd\gamma). \nonumber \ea
The fiducial co-triads determine a fiducial 3-metric $\q_{ab}
:= \w_a^i\w_{bi}$,
\be \q_{ab} \dd x^a \dd x^b =  \dd\alpha^2\,+\,\dd\beta^2\,+\,
\dd\gamma^2\,+\,2\cos\beta\dd\alpha\dd\gamma. \ee
It follows that $\sqrt{\q}=\sin\beta$ and one can see that $\q_{ab}$
is the metric of a 3-sphere with a volume $V_o = 16\pi^2$, this agrees with
$V_o = 2 \pi^2 r_o^3$ for $r_o = 2$ as specified above.  Finally, we
introduce the length-scale $\ell_o = V_o^{1/3}$ for later convenience.

In diagonal Bianchi models, the physical triads $e^a_i$ are related
to the fiducial ones by%
\footnote{There is no sum if repeated indices are both covariant or
contravariant.  As usual, the Einstein summation convention holds if a
covariant index is contracted with a contravariant index.}
\be \label{edef} \omega_a^i = a^i(\tau)\w_a^i \qquad \mathrm{and}
\qquad a_i(\tau) e^a_i = \e^a_i, \ee
where the $a_i$ are the three directional scale factors.

For later use, let us calculate the spin connection determined by
physical triads $e^a_i$.  Since $\Gamma_a^i$ is given by
\be \Gamma_a^i = -\epsilon^{ijk}\,e^b_j\,\left(\partial_{[a}
\omega_{b]k}+\f{1}{2}e^c_k \omega^l_a\partial_{[c}\omega_{b]l}
\right)\, , \ee
it follows that
\be \label{spin1} \Gamma_a^1 = \f{\ve}{r_o}\left(\f{a_1^2}{a_2a_3}
- \f{a_2}{a_3} - \f{a_3}{a_2}\right) \w_a^1, \ee
where $\ve := \epsilon_{123}$ is $+1$ for right-handed physical
triads and $-1$ for left-handed physical triads.  Note that while
the $\epz_{ijk}$ appearing in the Bianchi IX structure constants are
\emph{not} affected by the handedness of the physical triads,
$\epsilon_{ijk}$ and $\ve$ on the other hand \emph{are} affected
by the handedness of $e^a_i$.  $\Gamma_a^2$ and $\Gamma_a^3$ can
be obtained by permutations of Eq.~(\ref{spin1}).

As is usual in LQC, we will now use the fiducial triads and
co-triads in order to introduce a convenient parametrization of
the phase space variables $E^a_i$ and $A_a^i$.  Because we have
restricted ourselves to the diagonal model and these fields are
symmetric under the Bianchi IX group, from each equivalence class
of gauge related phase space variables we can choose a pair of
the form
\be \label{var} E^a_i = \f{p_i}{\ell_o^2}\sqrt{\q}\:\e^a_i \qquad {\rm
and} \qquad A_a^i = \f{c^i}{\ell_o}\w_a^i, \ee
where, as spelled out in footnote 3, there is no sum over $i$. Note
that the length $\ell_o$ plays a similar role to that of the lengths
of the fiducial cell in noncompact space-times in terms of the form
of the basic variables $(A, E)$.  In this case the manifold is compact
and there is no fiducial cell.

It is straightforward to relate the scale factors $a_i$ to the $p_i$:
\be p_1 = \sgn(a_1)|a_2a_3|\ell_o^2\, , \qquad p_2 = \sgn(a_2)|a_1a_3|\ell_o^2
\, , \qquad p_3 = \sgn(a_3)|a_1a_2|\ell_o^2\,, \ee
it follows that $\sqrt{|q|} = \sqrt{|p_1p_2p_3|}V_o^{-1}\sqrt{\q}$.

Thus, a point in the phase space is now coordinatized by six real
numbers $(p_i, c^i)$. One can use the symplectic structure in full
general relativity to induce a symplectic structure on the
six-dimensional phase space. The non-zero Poisson brackets are
given by
\be \label{pb} \{c_i,\, p_j\} \, = \, 8\pi G \gamma \,\delta_{ij}
\, , \ee
where $\gamma$ is the Barbero-Immirzi parameter.

Our choice (\ref{var}) of physical triads and connections has fixed
the internal gauge as well as the diffeomorphism freedom.
Furthermore, it is easy to explicitly verify that the Gauss and the
diffeomorphism constraints are automatically satisfied due to
Eq.~(\ref{var}). Thus we are left with the Hamiltonian constraint
\be \label{Hgen} \mathcal{C}_H = \int_\mathcal{M}
\Big[\f{-NE^a_iE^b_j}{16\pi G\gamma^2\sqrt{|q|}}\epsilon^{ij}{}_k
\Big(F_{ab}{}^k-(1+\gamma^2)\Omega_{ab}{}^k \Big)
+ N \mathcal{H}_{\m}\Big]\, \dd^3x\, \approx 0 , \ee
where $F_{ab}{}^k$ and $\Omega_{ab}{}^k$ are the curvature of $A_a^i$
and $\Gamma_a^i$ respectively, while $\mathcal{H}_\m$ is the matter
Hamiltonian density.  The $\approx 0$ indicates that $\mathcal{C}_H$
is a constraint and must vanish for physical solutions.  Since we
are most interested in the gravitational sector, our matter
field will consist only of a massless scalar field $T$ which will
later serve as a relational time variable \`a la Liebniz.
(Additional matter fields can be incorporated in a straightforward
manner, modulo possible intricacies of essential self-adjointness.)
Thus,
\be \mathcal{H}_{\m} = \f{1}{2}\f{p_T^2}{\sqrt{|q|}}. \ee

Since we want to use the massless scalar field as relational time,
it is convenient to use a harmonic-time gauge, i.e., assume that the
time coordinate $\tau$ satisfies $\Box \tau=0$.
The corresponding lapse function is $N=\sqrt{|p_1p_2p_3|}$. With
this choice, the Hamiltonian constraint simplifies considerably.

In terms of $p_i$, the first component of the spin connection is
given by
\be \Gamma_a^1 = \f{\ve}{r_o}\left(\f{p_2p_3}{p_1^2} - \f{p_2}{p_3}
- \f{p_3}{p_2}\right) \w_a^1, \ee
the other two spin connection components can be obtained via permutations.
The curvature of $\Gamma_a^i$ is in turn
\ba \Omega_{ab}{}^1 &=& 2 \partial_{[a}\Gamma_{b]}^1 + \epsilon^1{}_{jk}
\Gamma_a^j\Gamma_b^k \nonumber \\ &=& \f{2\ve}{r_o^2}\left(3\f{p_2p_3}{p_1^2}
+2\f{p_1^2}{p_2p_3}-2\f{p_2}{p_3}-2\f{p_3}{p_2}-\f{p_1^2p_2}{p_3^3}
-\f{p_1^2p_3}{p_2^3}\right)\w_{[a}^2\w_{b]}^3, \ea
the other components of $\Omega_{ab}{}^k$ can again be obtained via
permutations.

Finally, it is straightforward to calculate the curvature of $A_a^i$.
For example,
\ba F_{ab}{}^1 &=& 2 \partial_{[a}A_{b]}^1 + \epsilon^1{}_{jk}
A_a^j A_b^k \nonumber \\ &=& 2\left( \f{2c_1}{\ell_or_o} +
\f{\ve c_2c_3}{\ell_o^2} \right)\w_{[a}^2\w_{b]}^3.  \ea
Using these results, one finds that the Hamiltonian constraint
(\ref{Hgen}) is given by
\ba \label{Hcl} \mathcal{C}_H&=&-\f{1}{8\pi G\gamma^2}
\Bigg(p_1p_2c_1c_2+p_2p_3 c_2c_3+p_3p_1c_3c_1 + \f{2\ell_o\ve}{r_o}
\big(p_1p_2c_3 +p_2p_3c_1+p_3p_1c_2\big) \nonumber \\
&& \qquad+\f{\ell_o^2}{r_o^2}(1+ \gamma^2)\,\bigg[2p_1^2+2p_2^2+2p_3^2-
\Big(\f{p_1p_2}{p_3}\Big)^2-\Big(\f{p_2p_3}{p_1}\Big)^2-\Big(
\f{p_3p_1}{p_2}\Big)^2\bigg]\Bigg) \nonumber \\
&& \qquad + \f{1}{2}p_T^2 \approx 0. \ea
Note that the constraint for the closed isotropic case is recovered
for $p_1=p_2=p_3$ while the Bianchi I constraint is recovered in the
limit $r_o\to\infty$ or, equivalently, $\ell_o\to0$.  We will take
advantage of this correspondence and set $r_o=2$ for the remainder of
the paper.  The Bianchi I limit can be obtained by taking $\ell_o\to0$.

One can now derive the time evolution of any classical observable
$\mathcal{O}$ by taking its Poisson bracket with $\mathcal{C}_H$:
\be \dot{\mathcal{O}} = \{\mathcal{O},\mathcal{C}_H\}\, , \ee
where the `dot' stands for derivative with respect to the harmonic time
$\tau$. This gives
\be \label{ceom1} \dot{p_1}=\f{p_1}{\gamma}\left(p_2c_2 + p_3c_3
+ \ell_o\ve \f{p_2p_3}{p_1}\right), \ee
\be \label{ceom2} \dot{c_1} =- \f{1}{\gamma} \Bigg(p_2c_1c_2 +
p_3c_1c_3 + \ell_o\ve(p_2c_3 + p_3c_2) + \ell_o^2 (1+\gamma^2) \bigg(p_1
+ \f{p_2^2p_3^2}{2p_1^3} - \f{p_1p_2^2}{2p_3^2}
- \f{p_1p_3^2}{2p_3^2}\bigg)\Bigg). \ee
As usual, the other equations of motion can be obtained by permutations.
Any initial data satisfying the Hamiltonian constraint can be evolved by
these equations of motion.  It is particularly interesting to study the
Hubble rates $H_i$ which are given by
\be H_i = \f{1}{a_i}\f{\d a_i}{\d t}, \ee
where $t$ is the proper time and is related to the harmonic time $\tau$
(which is the time coordinate used until now) by
\be \f{\d}{\d t} = \f{1}{\sqrt{|p_1p_2p_3|}}\f{\d}{\d\tau}. \ee
It follows that the Hubble rates are related to the $(c_i, p_i)$ by, e.g.,
\be \label{Hiclass} c_1 p_1 = \gamma \sqrt{|p_1 p_2 p_3|} H_1 + \f{\ell_o}{2}
\left( \f{p_2 p_3}{p_1} - \f{p_1 p_2}{p_3} - \f{p_1 p_3}{p_2} \right). \ee
The mean Hubble rate $H$ of the mean scale factor $a = (a_1a_2a_3)^{1/3}$
is given by
\be H = \f{1}{a}\f{\d a}{\d t} = \f{1}{3}(H_1 + H_2 + H_3), \ee
and the Friedmann equation is
\be \label{H2class} H^2 = \f{8\pi G}{3}\rho + \f{1}{6}\sigma^2
- \f{\ell_o^2}{12}V(p), \ee
where the energy density of the scalar field is $\rho = p_T^2/2|p_1 p_2 p_3|$,
the shear term is given by
\be \sigma^2 = \f{1}{3}[(H_1 - H_2)^2 + (H_2 - H_3)^2 + (H_3 - H_1)^2], \ee
and the potential is
\be V(p) = \f{1}{p_1p_2p_3} \left[2(p_1^2 + p_2^2 + p_3^2)
- \left(\f{p_2p_3}{p_1}\right)^2 - \left(\f{p_3p_1}{p_2}\right)^2
- \left(\f{p_1p_2}{p_3}\right)^2 \right]. \ee
Clearly, these dynamics are quite complex already at the classical level
and, as mentioned in the introduction, become chaotic as a singularity is
approached.  The one exception is the case when the matter field is a
massless scalar field which is precisely what is considered here.  In
this case, as the singularity is approached, the Friedmann equation is
asymptotically velocity term dominated (AVTD) which means that the
potential can be safely neglected \cite{bb}.  Thus, as the singularity is
approached, the dynamics are the same as those of the Bianchi I space-time
with a massless scalar field.  This behaviour will be important for the
study of the effective equations later.  However, the quantum Hamiltonian
constraint operator derived in the following section will hold everywhere
and it will be relatively straightforward to extend it for other types of
matter fields which classically allow the full Mixmaster dynamics.

Finally, before moving on to the quantum theory, let us consider the
parity transformation $\Pi_k$ which flips the $k$th \emph{physical}
triad vector $e^a_k$. (Keep in mind that this transformation does
not act on any of the fiducial quantities which carry the label $o$.)
These correspond to residual discrete gauge transformations.  Under
this map, we have: $q_{ab} \to q_{ab}, \, \epsilon_{abc} \to
\epsilon_{abc}\,$ but $\epsilon_{ijk} \to -\epsilon_{ijk}, \, \ve
\to -\ve$.  The canonical variables $c_i, p_i$ transform as proper
internal vectors and co-vectors.  For example,
\be \Pi_1(c_1,c_2,c_3) \rightarrow (-c_1, c_2, c_3) \qquad {\rm and}
\qquad \Pi_1(p_1,p_2,p_3) \rightarrow (-p_1, p_2, p_3)\, . \ee
Consequently, both the symplectic structure and the Hamiltonian
constraint are left invariant under any of the parity maps $\Pi_k$.

The Hamiltonian description given in this section will serve as the
starting point for the loop quantization in the next section.

\section{Quantum Theory}
\label{s3}

This section is divided into three parts.  In the first, we discuss
the kinematics of the model and in the second we introduce the
Hamiltonian constraint operator and describe its action on
physical states.  Finally, in the third subsection we show that the
dynamics of a wave function sharply peaked around an isotropic
geometry are well approximated by the LQC dynamics of the closed
FRW model.

\subsection{LQC Kinematics}
\label{s3.1}

The kinematics for the LQC of Bianchi IX models is identical to that
of the Bianchi II models \cite{awe3}, but we will briefly present the
kinematics here as well for the sake of completeness.

The elementary functions on the classical
phase space that have unambiguous analogs in the quantum
theory are the momenta $p_i$ and holonomies $h_k^{(\mu)}$ of the
gravitational connection $A_a^i$ along the integral curves of
$\e^a_k$ of length $\mu \ell_o$ with respect to the fiducial metric
$\q_{ab}$. These holonomies are given by
\be \label{hol} h_k^{(\mu)}(c_1,c_2,c_3) = \exp\left(\mu
c_k\tau_k\right) = \cos\f{\mu c_k}{2} \mathbb{I} + 2\sin\f{\mu
c_k}{2}\tau_k, \ee
where the $\tau_k$ are $-i/2$ times the Pauli matrices.  This family
of holonomies is completely determined by the almost periodic functions
$\exp(i\mu c_k)$ of the connection. These almost periodic functions
will be the elementary configuration variables which will be promoted
unambiguously to operators in the quantum theory.

It is simplest to use the $p$-representation to  specify the
gravitational sector $\Hkg$ of the kinematic Hilbert space. The
basis is orthonormal in the sense that
\be \langle p_1,p_2,p_3|p_1',p_2',p_3'\rangle = \delta_{p_1^{}p_1'}
\delta_{p_2^{}p_2'}\delta_{p_3^{}p_3'}\, , \ee
where the right side features Kronecker delta symbols rather than
Dirac delta distributions. Kinematical states consist of
countable linear combinations
\be |\Psi\rangle \,=\,
\sum_{p_1,p_2,p_3}\Psi(p_1,p_2,p_3)|p_1,p_2,p_3\rangle\ \ee
of these basis states for which the norm
\be \label{norm} ||\Psi ||^2\, =\, \sum_{p_1,p_2,p_3}\,
|\Psi(p_1,p_2,p_3)|^2 \ee
is finite.

Next, recall that on the classical phase space the three reflections
$\Pi_i:\,\,e^a_i\,\to\, -e^a_i$ are large gauge transformations
under which physics does not change since both the metric and the
extrinsic curvature are left invariant. These large gauge
transformations have a natural induced action, denoted by
$\hat\Pi_i$, on the space of wave functions $\Psi(p_1,p_2,p_3)$. For
example,
\be \hat\Pi_1\Psi(p_1,p_2,p_3)=\Psi(-p_1,p_2,p_3). \ee
Since $\hat\Pi_i^2$ is the identity, for each $i$ the group of
these large gauge transformations is simply $\Z_2$. As in Yang-Mills
theory, physical states belong to its irreducible representation.  For
definiteness, as in the isotropic and the Bianchi type I and type II
models, we will work with the symmetric representation. It then follows
that $\mathcal{H}_{\mathrm{kin}}^{\mathrm{grav}}$ is spanned by wave
functions $\Psi(p_1,p_2,p_3)$ which satisfy
\be \label{parity} \Psi(p_1,p_2,p_3)=\Psi(|p_1|,|p_2|,|p_3|) \ee
and have a finite norm.

The action of the elementary operators on
$\mathcal{H}_{\mathrm{kin}}^{\mathrm{grav}}$ is as follows: the
momenta act by multiplication whereas the almost periodic functions
in $c_i$ shift the $i$th argument. For example,
\be [\hat p_1 \Psi](p_1,p_2,p_3) = p_1\, \Psi(p_1,p_2,p_3) \,\quad
\mathrm{and} \,\quad \Big[\widehat{\exp(i\mu c_1)}\Psi\Big](p_1,
p_2, p_3) = \Psi(p_1-8\pi\gamma G\hbar \mu, p_2, p_3)\, . \ee
The expressions for $\hat p_2, \widehat{\exp(i\mu c_2)}, \hat p_3$
and $\widehat{\exp(i\mu c_3)}$ are analogous.  Finally, we must
define the operator $\hat{\ve}$ since $\ve$ features in the
expression of the Hamiltonian constraint.  Following \cite{awe3},
we define
\be \label{ve2} \hat{\ve}\,|p_1,p_2,p_3\rangle := \left\{
\rlap{\raise2ex\hbox{\,\,$\quad|p_1,p_2,p_3 \rangle$ if $p_1p_2p_3
\ge 0$,}}{\lower2ex\hbox{\,\,$ -\,|p_1,p_2,p_3 \rangle$ if
$p_1p_2p_3<0$.}} \right. \ee
Finally, the full kinematical Hilbert space
$\mathcal{H}_{\mathrm{kin}}$ will be the tensor product
$\mathcal{H}_{\mathrm{kin}}=\mathcal{H}_{\mathrm{kin}}^
{\mathrm{grav}}\otimes\mathcal{H}_{\mathrm{kin}}^{\mathrm{matt}}$,
where $\mathcal{H}_{\mathrm{kin}}^{\mathrm{matt}}=L^2({\R},dT)$ is
the matter kinematical Hilbert space for the homogeneous scalar
field.  On $\mathcal{H}_ {\mathrm{kin}}^{\mathrm{matt}}$, $\hat T$
will act by multiplication and $\hat p_T:=-i\hbar \mathrm{d}_T$ will
act by differentiation.

\subsection{The Quantum Hamiltonian Constraint}
\label{s3.2}

To define the quantum Hamiltonian constraint, we must express
the Hamiltonian constraint in terms of almost periodic functions
of the connection which can be directly promoted to operators.
For isotropic and/or spatially flat space-times, this can be
done by expressing the field strength $F_{ab}{}^k$ in terms of
holonomies and this is what is done for the $\bar\mu_i$ approach
in LQC in \cite{aps3, apsv, warsaw1, awe2}.  However, this is not
possible for space-times which are both anisotropic and spatially
curved such as the Bianchi type II and type IX models.  In this
case we need to extend the strategy: the connection itself ---rather
than the field strength--- has to be expressed in terms of
holonomies.  This task was carried out in \cite{awe3}.  The
connection operator is given by
\be \label{c-op} \hat{c}_k = \f{\widehat{\sin(\bar\mu_kc_k)}}{\bar\mu_k}, \ee
where
\be \label{mubar} \bar\mu_1 = \sqrt\f{|p_1|\Delta\,\lp^2}{|p_2p_3|},
\qquad  \bar\mu_2 = \sqrt\f{|p_2|\Delta\,\lp^2}{|p_1p_3|}, \qquad
\bar\mu_3 = \sqrt\f{|p_3| \Delta\,\lp^2}{|p_1p_2|}, \ee
and $\Delta\,\lp^2 = 4\sqrt{3}\pi\gamma\,\lp^2$ is the `area gap'.
Note that the choice for this operator is motivated by LQG: it is
obtained in \cite{awe3} by expressing the connection in terms of
holonomies, a procedure commonly used in LQG, and then ensuring
that this approach is equivalent to what is done for simpler
cosmological models.  Although the precise value of the area gap
may change as the relation between LQG and LQC is better
understood, the form of $\bar\mu_i$ in terms of the $p_i$ is necessary
in order to obtain the correct infrared, low curvature behaviour.

Using the connection operator, it is possible to promote the
classical Hamiltonian constraint in Eq.~(\ref{Hcl}) to an
operator.  Ignoring factor ordering ambiguities and inverse triad
operators for the moment, $\hat{\mathcal{C}}_H$ is given by
\begin{align} \label{qHam1} \hat{\mathcal{C}}_H =& -\f{1}{8\pi
G\gamma^2\Delta\lp^2}\Bigg[p_1 p_2|p_3|\sin\bar\mu_1c_1
\sin\bar\mu_2c_2+|p_1|p_2p_3\sin\bar\mu_2c_2\sin\bar\mu_3c_3 \nonumber \\
&+p_1|p_2|p_3\sin\bar\mu_3c_3\sin\bar\mu_1c_1\Bigg]-
\f{\ell_o\hat{\ve}}{8\pi G\gamma^2\sqrt{\Delta}\lp}\Bigg[
p_1p_2\sqrt\f{|p_1p_2|}{|p_3|}\sin\bar\mu_3c_3 \nonumber \\ &+
p_2p_3\sqrt\f{|p_2p_3|}{|p_1|}\sin\bar\mu_1c_1 +
p_3p_1\sqrt\f{|p_3p_1|}{|p_2|}\sin\bar\mu_2c_2\Bigg]
-\f{\ell_o^2(1+\gamma^2)}{32\pi G\gamma^2}\Bigg[
2\left(p_1^2+p_2^2+p_3^2\right)\nonumber \\ & 
-\left(\f{p_1p_2}{p_3}\right)^2
-\left(\f{p_2p_3}{p_1}\right)^2
-\left(\f{p_3p_1}{p_2}\right)^2
\Bigg]+\f{1}{2}\hat{p}_T^2, \end{align}
where for simplicity of notation here and in what follows we have
dropped the hats on the $p_i$ and $\sin\bar\mu_ic_i$ operators.

To obtain the action of the $\sin\bar\mu_ic_i$ operators (or,
equivalently, the $\exp(i\bar\mu_ic_i)$ operators) we will use the
same strategy as in \cite{awe2}.  As shown there, it is simplest
to introduce the dimensionless variables
\be \la_i=\f{\sgn(p_i)\sqrt{|p_i|}}{(4\pi\gamma\sqrt\Delta
\lp^3)^{1/3}}\, . \ee
Then the kets $|\la_1,\la_2,\la_3\rangle$ constitute an orthonormal
basis in which the operators $p_k$ are diagonal
\be p_k|\la_1,\la_2,\la_3\rangle\, =\,
[\sgn(\la_k)(4\pi\gamma\sqrt\Delta\lp^3)^{2/3}
\la_k^2]\,\,|\la_1,\la_2,\la_3\rangle\, , \ee
and quantum states are represented by functions $\Psi(\la_1,\la_2,\la_3)$.
Then the operator $e^{i\bar\mu_1c_1}$ acts by shifting the wavefunction,
\begin{align} \big[e^{i\bar\mu_1c_1}\,\Psi\big] (\la_1,\la_2,\la_3)
&= \Psi(\la_1- \f{1}{|\la_2\la_3|},\la_2,\la_3) \nonumber \\
&= \Psi(\f{v-2\sgn(\la_2\la_3)}{v}\cdot\, \la_1,\la_2,\la_3),
\end{align}
where we have introduced the variable $v=2\la_1\la_2\la_3$ which is
proportional to the volume V of the space-time:
\be \hat{V}\,\Psi(\la_1,\la_2,\la_3)\, =\,
[2\pi\gamma\sqrt\Delta\,|v|\,\lp^3]\, \Psi(\la_1, \la_2,\la_3). \ee
The action of the operators $e^{i\bar\mu_2 c_2}$ and $e^{i\bar\mu_3c_3}$
is analogous.

We are now ready to write the Hamiltonian constraint explicitly in
the $\la_i$-representation, again ignoring factor-ordering issues for
the time being:
\be \label{qHam2} \hat{\mathcal{C}}_H = \hat{\mathcal{C}}_1
+\hat{\mathcal{C}}_2+\hat{\mathcal{C}}_3+\hat{\mathcal{C}}_4+\tf{1}{2}
\hat{p}_T^2, \ee
where
\begin{align} \hat{\mathcal{C}}_1 &= -\tf{1}{2}\pi\hbar\lp^2v^2\Big[
\sgn(\la_1\la_2)\sin\bar\mu_1c_1\sin\bar\mu_2c_2+\sgn(\la_2\la_3)
\sin\bar\mu_2c_2\sin\bar\mu_3c_3 \nonumber \\ & \quad\qquad +
\sgn(\la_3\la_1)\sin\bar\mu_3c_3\sin\bar\mu_1c_1 \Big]; \\
\hat{\mathcal{C}}_2 &= -2\pi\sqrt{\Delta}\hbar\lp^3\ell_o\hat{\ve}\Big[
(\la_1\la_2)^3\f{1}{\sqrt{|p_3|}}\sin\bar\mu_3c_3+ (\la_2\la_3)^3
\f{1}{\sqrt{|p_1|}}\sin\bar\mu_1c_1 \nonumber \\ & \quad\qquad +
(\la_3\la_1)^3\f{1}{\sqrt{|p_2|}}\sin\bar\mu_2c_2\Big]; \\
\hat{\mathcal{C}}_3 &= -\f{(4\pi\gamma\sqrt{\Delta})^{1/3}\sqrt{\Delta}
\hbar\lp^2}{4\gamma}\ell_o^2(1+\gamma^2)\Big[\la_1^4+\la_2^4+\la_3^4\Big]; \\
\hat{\mathcal{C}}_4 &= \tf{1}{2}(16\pi^2\gamma^2\Delta)^{1/3}\pi
\Delta\hbar\lp^6\ell_o^2(1+\gamma^2)\Big[(\la_1\la_2)^4\f{1}{p_3^2}+
(\la_2\la_3)^4\f{1}{p_1^2}+(\la_3\la_1)^4\f{1}{p_2^2}\Big].
\end{align}

It will be straightforward to deal with $\hat{\mathcal{C}}_H$ since the
terms in $\hat{\mathcal{C}}_1$ are the exact terms that appear in the
Bianchi I model and have already been studied in \cite{awe2} while the
terms in $\hat{\mathcal{C}}_2$ and $\hat{\mathcal{C}}_4$ are of the same
form as some of the terms in the Bianchi II model \cite{awe3}.
Finally, the only new terms ---those in $\hat{\mathcal{C}}_3$--- act by
multiplication and will not cause any difficulty.

All of the terms will be factor-ordered in a symmetric manner.
For example, the first term in $\hat{\mathcal{C}}_1$ will be
factor-ordered as
\begin{align} \label{factorord} -\f{1}{16}\pi\hbar\lp^2\sqrt{|v|}
\Bigg[(\sin\bar\mu_1c_1 \sgn\la_1 + \sgn\la_1 \sin\bar\mu_1c_1) |v|
(\sin\bar\mu_2c_2 \sgn\la_2 + \sgn\la_2 \sin\bar\mu_2c_2) \nonumber \\
+ (\sin\bar\mu_2c_2 \sgn\la_2 + \sgn\la_2 \sin\bar\mu_2c_2) |v|
(\sin\bar\mu_1c_1 \sgn\la_1 + \sgn\la_1 \sin\bar\mu_1c_1)\Bigg]
\sqrt{|v|}, \end{align}
while the first term in $\hat{\mathcal{C}}_2$ will be
\be -\pi\sqrt{\Delta}\hbar\lp^3\ell_o(\la_1\la_2)^3\f{1}{|p_3|^{1/4}}\Big[
\hat{\ve}\sin\bar\mu_3c_3+\sin\bar\mu_3c_3\hat{\ve}\Big]
\f{1}{|p_3|^{1/4}}. \ee
Since all of the components in each term in $\hat{\mathcal{C}}_3$
and $\hat{\mathcal{C}}_4$ commute, there are no factor-ordering choices
to be made for these terms.

The factor ordering given in Eq.~(\ref{factorord}) was first introduced
in \cite{gowdy} for the study of the Gowdy model.  It is a particularly
nice choice as it causes the octants to decouple from each other, one can
then focus on the dynamics of a single octant and then derive the behaviour
of the other octants via the parity properties of the wave function.

The only operators that remain to be defined are the inverse volume
operators.  Using a variation on the Thiemann inverse triad identities
\cite{tt}, one obtains the operator \cite{awe3}
\be \label{inv} \widehat{|p_1|^{-1/4}}\, |\la_1,\la_2,\la_3\rangle =
\f{\sqrt2 \sgn(\la_1)\,\sqrt{|\la_2\la_3|}}
{(4\pi\gamma\sqrt\Delta\lp^3)^{1/6}}
\left(\sqrt{|v+\sgn(\la_2\la_3)|}-\sqrt{|v-\sgn(\la_2\la_3)|}\right)\,
|\la_1,\la_2,\la_3\rangle. \ee
This operator is diagonal in the eigenbasis $|\la_1,\la_2,\la_3\rangle$
and, on eigenkets with large volume, the eigenvalue is indeed
well approximated by $|p_1|^{-1/4}$, whence on semi-classical states
it behaves as the inverse of $|\hat{p}|^{1/4}$, just as one would
hope.  Nonetheless, there are interesting nontrivialities
in the Planck regime, the most important one being that
the inverse triad operator annihilates states
$|\la_1,\la_2,\la_3\rangle$ where $v = 2\la_1\la_2\la_3 = 0$.

Finally, the other inverse triad operator which is necessary for the
study of Bianchi IX models can be defined by
\be \widehat{p_i^{-2}} := \left(\widehat{|p_i|^{-1/4}}\right)^8. \ee
Note that both of these operators were already introduced for the
study of the Bianchi II model in \cite{awe3}.

As in the Bianchi I model, the action simplifies if we replace
$(\la_i, \la_j, \la_k)$ by $(\la_i, \la_j, v)$%
\footnote{This cannot be done for states where $\la_1\la_2\la_3 = 0$
but since these states decouple under the action of
$\hat{\mathcal{C}}_H$, we can restrict our attention solely to states
where $\la_1\la_2\la_3 \ne 0$.}.
Because of the high symmetry of the Bianchi IX model, it does not
matter which of the $\la_i$ is replaced; we will choose to replace
$\la_3$ by $v$ here.  This change of variables would be nontrivial
if, as in the Wheeler-DeWitt theory, we had used the Lesbegue
measure in the gravitational sector. However, it is quite tame
here because the norms are defined using a discrete measure.  The
inner product on $\Hkg$ is now given by
\be \langle\Psi_1|\Psi_2\rangle_{\rm kin} = \sum_{\la_1,\la_2,v}
\,\,\bar{\Psi}_1(\la_1,\la_2,v)\,\Psi_2(\la_1,\la_2,v) \ee
and states are symmetric under the action of $\hat\Pi_k$. In the
Appendix of \cite{awe3}, it is shown that under the action of the
$\hat\Pi_i$, the operators $\sin\bar\mu_ic_i$ have the same
transformation properties as $c_i$ under the reflections $\Pi_i$
in the classical theory. As a consequence, $\hat{\mathcal{C}}_H$ is
also reflection symmetric%
\footnote{Note that although $\hat\Pi_i\hat{\ve}\hat\Pi_i = -\hat{\ve}$
(recall that classically $\ve\to-\ve$ under a parity transformation)
only when $v\ne0$, in the $v=0$ case the wavefunction is annihilated
by the gravitational part of the Hamiltonian constraint
$\hat{\mathcal{C}}_\g$ and therefore $\hat\Pi_i\hat{\mathcal{C}}_\g
\hat\Pi_i|\Psi_{\rm sing}\rangle = 0 = \hat{\mathcal{C}}_\g
|\Psi_{\rm sing}\rangle$ where $|\Psi_{\rm sing}\rangle$ is a state that
only has support on $v=0$.  It is then straightforward to
show that $\hat\Pi_i\hat{\mathcal{C}}_H\hat\Pi_i|\Psi\rangle
= \hat{\mathcal{C}}_H|\Psi\rangle$ for all wavefunctions.}%
. Therefore, its action is well defined on
$\Hkg$: $\hat{\mathcal{C}}_H$ is a densely defined, symmetric
operator on this Hilbert space. In the isotropic and Bianchi I cases,
its analog has been shown to be essentially self-adjoint \cite{warsaw2, kp}.
In what follows we will assume that (\ref{qHam2}) is essentially
self-adjoint on $\Hkg$ and work with its self-adjoint extension.

We can now study the action of $\hat{\mathcal{C}}_H$ on a wavefunction.
For a complete derivation of the action of each term in the constraint,
see \cite{awe2, awe3}.

It is straightforward to write down the full Hamiltonian
constraint on $\Hkg$:
\be \label{qHam4} -\hbar^2\, \partial^2_T \, \Psi(\la_2,\la_3,v; T) =
\Th\, \Psi(\la_2,\la_3,v; T), \quad {\rm where}\quad \Th = -2
\hat{\mathcal{C}}_\g. \ee
As in the isotropic case \cite{aps2}, one can obtain the physical
Hilbert space $\H_{\rm phy}$ by a group averaging procedure and the
final result is completely analogous. Elements of $\H_{\rm phy}$
consist of `positive frequency' solutions to (\ref{qHam4}), i.e.,
solutions to
\be \label{qHam5} -i\hbar \partial_T \Psi(\la_1,\la_2,v; T)\, = \,
\sqrt{|\Th|}\, \Psi(\la_1,\la_2,v; T)\, ,\ee
which are symmetric under the three reflection maps $\hat\Pi_i$:
\be \label{sym} \Psi(\la_1,\la_2,v;\, T) = \Psi(|\la_1|,|\la_2|,|v|;\,
T)\, . \ee
The scalar product is simply given by
\ba \label{ip1} \langle \Psi_1|\Psi_2\rangle_{\rm phys} &=& \langle
\Psi_1(\la_1,\la_2,v; T_o)|\Psi_2(\la_1,\la_2,v; T_o) \rangle_{\rm
kin} \nonumber\\
&=& \sum_{\la_1,\la_2,v} \bar\Psi_1(\la_1,\la_2,v; T_o)\,
\Psi_2(\la_1,\la_2,v; T_o), \ea
where $T_o$ is any ``instant'' of internal time $T$.

Since elements of $\Hkg$ are invariant under the three parity maps
$\hat{\Pi}_k$ and the Hamiltonian constraint satisfies $\hat{\Pi}_k\,
\hat{\mathcal{C}}_\g \hat{\Pi}_k = \hat{\mathcal{C}}_\g$, knowledge
of the restriction of the image $\hat{ \mathcal{C}}_\g\Psi$ of $\Psi$
to the positive octant suffices to determine $\hat{\mathcal{C}}_\g
\Psi$ completely.  Therefore, in the remainder of this section we will
restrict the argument of $\hat{\mathcal{C}}_H\Psi$ to the positive
octant. The full action is simply given by
\be \big(\hat{\mathcal{C}}_\g\Psi\big)(\la_1,\la_2, v)=\big(\hat{
\mathcal{C}}_\g\Psi\big)(|\la_1|,|\la_2|,|v|). \ee

Since all states with $v=0$ are annihilated by $\hat{\mathcal{C}}_\g$,
their evolution is trivial:
\be \label{qHamsing} \partial_T^2\, \Psi(\la_1,\la_2,v=0;T) = 0\, . \ee
Such states correspond to classical geometries which are singular and
therefore we will call these states `singular', even though they are
well defined in the quantum theory.  Non-singular states on the other
hand are physically much more interesting. On them, the explicit form
of the full constraint is given by:
\begin{align} \partial_T^2\, \Psi(\la_1,\la_2,v;T) =& \pi G\Bigg[\f{\sqrt{v}}{8}
\bigg((v+2)\sqrt{v+4}\,\Psi^+_4(\la_1,\la_2,v;T) - (v+2)\sqrt v\,
\Psi^+_0( \la_1, \la_2,v;T)\nonumber \\& -\theta_{v-2} (v-2)\sqrt v\,
\Psi^-_0(\la_1,\la_2,v;T)+\theta_{v-4}(v-2)\sqrt{|v-4|}
\,\Psi^-_4(\la_1,\la_2,v;T)\bigg)\nonumber \\ &
-\f{2i\ell_o\sqrt\Delta}{(4\pi \gamma\sqrt\Delta)^{1/3}}
\left(\sqrt{v+1}-\sqrt{|v-1|}\right)\bigg(
\Phi^+-\theta_{v-2}\Phi^-\bigg)(\la_1,\la_2,v;T) \nonumber\\&
+\f{8\Delta \ell_o^2(1+\gamma^2)}{(4\pi\gamma\sqrt\Delta)^{2/3}}\bigg(
(\sqrt{v+1}-\sqrt{|v-1|})^8 \Big[(\la_1\la_2)^8+(\la_2\la_3)^8
+(\la_3\la_1)^8\Big]  \nonumber \\ \label{qHamfin} &
-\f{1}{8}\big(\la_1^4+\la_2^4+\la_3^4\big) \bigg)
\:\Psi(\la_1,\la_2,v;T)\Bigg], \end{align}
where $\theta_x$ is the step function
\be \theta_x = \left\{ \rlap{\raise2ex\hbox{\,\,1 if $x > 0$,}}
{\lower2ex\hbox{\,\,0 if $x < 0$.}} \right. \ee
Note that the step function kills any terms that would allow the positive octant
to interact with any of the other ones, this is a direct consequence of the
factor ordering choices made earlier.

The $\Psi^\pm_{0,4}$ are defined as follows:
\begin{align} \Psi^\pm_4(\la_1,\la_2,v;T)=& \:\Psi\left(\f{v\pm4}{v\pm2}\cdot
\la_1,\f{v\pm2}{v}\cdot\la_2,v\pm4;T\right)+\Psi\left(\f{v\pm4}{v\pm2}\cdot\la_1,
\la_2,v\pm4;T\right)\nonumber\\& +\Psi\left(\f{v\pm2}{v}\cdot\la_1,\f{v\pm4}{v
\pm2}\cdot\la_2,v\pm4;T\right)+\Psi\left(\f{v\pm2}{v}\cdot\la_1, \la_2,v\pm4;T
\right) \nonumber \\ & +\Psi\left(\la_1,\f{v\pm2}{v}\cdot\la_2,v\pm4;T\right)+
\Psi\left(\la_1,\f{v\pm4}{v\pm2}\cdot\la_2,v\pm4;T\right), \end{align}
and
\begin{align} \Psi^\pm_0(\la_1,\la_2,v;T)= & \:\Psi\left(\f{v\pm2}{v}\cdot\la_1,
\f{v}{v\pm2}\cdot\la_2,v;T\right)+\Psi\left(\f{v\pm2}{v}\cdot\la_1,\la_2,v;T
\right) \nonumber \\ & +\Psi\left(\f{v}{v\pm2}\cdot\la_1,\f{v\pm2}{v}\cdot\la_2,
v;T\right)+\Psi\left(\f{v}{v\pm2}\cdot\la_1,\la_2,v;T\right) \nonumber \\ & +
\Psi\left(\la_1,\f{v}{v\pm2}\cdot\la_2,v;T\right)+\Psi\left(\la_1,\f{v\pm2}{v}
\cdot\la_2,v;T\right)\, ,\end{align}
while $\Phi^\pm$ are given by
\begin{align} \Phi^\pm(\la_1,\la_2,v;T)\, =&
\big(\sqrt{|v\pm2+1|}-\sqrt{|v\pm2-1|}\big) \times \bigg[(\la_2\la_3)^4
\Psi\Big(\f{v\pm2}{v}\cdot\la_1,\la_2,v\pm2;T\Big) \nonumber \\ &
\label{defPhi}+(\la_3\la_1)^4\Psi\Big(\la_1,\f{v\pm2}{v}\cdot\la_2,v\pm2;T\Big)
+ (\la_1\la_2)^4 \Psi\Big(\la_1,\la_2,v\pm2;T\Big) \bigg]. \end{align}

As expected, the quantum dynamics of the Bianchi IX model reduces to
that of the Bianchi I model discussed in \cite{awe2} in the limit
$\ell_o\to0$ in Eq.~(\ref{qHamfin}).

Eq.~(\ref{qHamfin}) also immediately implies that the steps in $v$ are
uniform: the argument of the wave function only involves $v-4, v-2, v,
v+2$ and $v+4$. Thus, there is a superselection in $v$. For each
$\epsilon\in[0,2),$ we can introduce a lattice $\mathcal{L}_\epsilon$
of points $v=2n+\epsilon$.  Then the quantum evolution ---and the
action of the Dirac observables $\hat{p}_T$ and $\hat{V}|_{T}$ commonly
used in LQC--- preserves the subspaces $\mathcal{H}^\epsilon_{\mathrm{phy}}$
consisting of states with support in $v$ on $\mathcal{L}_\epsilon$.
The most interesting lattice is the one corresponding to
$\epsilon=0$ since it includes the classically singular points $v=0$.

The form of the action of the Hamailtonian constraint operator also shows
that the classical singularity is resolved.  Using the scalar field $T$
as time, we find that if one starts with a wavefunction which only has
support on singular states, that wavefunction does not evolve in $T$ and
therefore will always only have support on singular states.

On the other hand, a state which does not have any support on
the singular subspace will never have support on it.  Restricting our
argument to the positive octant for the sake of simplicity (it can easily
be generalized to the other octants), it is easy to see that to go from
$\la_1, \la_2, v > 0$ to $v=0$, one must either have $v=2$ and then $\Phi^-$
will give a term with $v=0$ or have $v=4$ and then $\Psi^-_4$ will give
a term with $v=0$.  However, the prefactors in front of $\Phi^-$ vanish
for $v=2$ just as the prefactors in front of $\Psi^-_4$ vanish for $v=4$.
Because of this, it is impossible for a wavefunction with no support on
singular states to ever gain support on a singular state.

This shows that singular states decouple from nonsingular states under
the relational $T$ dynamics given by Eqs.~(\ref{qHamsing}) and
(\ref{qHamfin}).  In other words, if one starts with a nonsingular state
at some `time' $T_o$, it will remain nonsingular throughout its evolution.
It is in this (rather strong) sense that the singularity is resolved.

\section{Effective Equations}
\label{s4}

In the isotropic models, effective equations have been introduced
via two different approaches ---the embedding \cite{jw, vt} and the
moment expansion \cite{bs} methods--- in order to study the first order
quantum-corrected equations of motion.  In the isotropic case the
effective equations following from the embedding approach provide
an excellent approximation to the full quantum evolution of states
which are Gaussians at late times, even in the $\Lambda\not=0$ as
well as k=$\pm 1$ cases where the models are not exactly soluble.
However, the truncation method is more systematic and also more
general in the sense that it is applicable to a wide variety of
states.  Nonetheless, in this section we will use the first method
(although we will ignore the effect of fluctutations in this work)
in order to gain qualitative insights into modifications of the
equations of motion due to quantum geometry effects.

To obtain the effective equations we can restrict our attention
to the positive octant of the classical phase space (where $\ve=1$)
without loss of generality. Then the quantum corrected Hamiltonian
constraint is given by the classical analogue of (\ref{qHam1}):
\be \label{Heff} \f{p_T^2}{2}+\mathcal{C}^{\mathrm{eff}}_{\mathrm{grav}}
= 0, \ee
where%
\footnote{Recall that every $\ell_o$ which appears in the constraint
is divided by $r_o$ which has been set to 2.  As $\ell_o/r_o$ is
dimensionless, we must ignore $\ell_o$ when counting units.}
\begin{align} \mathcal{C}^{\mathrm{eff}}_{\mathrm{grav}} =&
-\f{p_1p_2p_3}{8\pi G\gamma^2\Delta\lp^2}
\Bigg[\sin\bar\mu_1c_1\sin\bar\mu_2c_2+\sin\bar\mu_2c_2
\sin\bar\mu_3c_3+\sin\bar\mu_3c_3\sin\bar\mu_1c_1\Bigg] \nonumber
\\ & \quad - \f{\ell_o}{8\pi G\gamma^2\sqrt\Delta\lp}\Bigg[
\f{(p_1p_2)^{3/2}\!\!\!}{\sqrt{p_3}}\sin\bar\mu_3c_3+
\f{(p_2p_3)^{3/2}\!\!\!}{\sqrt{p_1}}\sin\bar\mu_1c_1+
\f{(p_3p_1)^{3/2}\!\!\!}{\sqrt{p_2}}\sin\bar\mu_2c_2\Bigg] \nonumber \\
& \quad -\f{\ell_o^2}{32\pi G \gamma^2}(1+\gamma^2)\Bigg[2(p_1^2+p_2^2+p_3^2) -
\left(\f{p_1p_2}{p_3}\right)^2
-\left(\f{p_2p_3}{p_1}\right)^2
-\left(\f{p_3p_1}{p_2}\right)^2
\Bigg]. \end{align}
Using the expressions (\ref{mubar}) of $\bar\mu_k$, it is easy to
verify that far away from the classical singularity ---more
precisely in the regime in which the Hubbles rates $H_i$ are well
below the Planck scale--- the effective Hamiltonian constraint
(\ref{Heff}) is well-approximated by the classical one given in
Eq.~(\ref{Hcl}).

The effective dynamics are obtained by taking Poisson brackets with
the effective Hamiltonian constraint.  This gives
\begin{align} \dot{p_1} &=
\gamma^{-1}\left(\f{p_1^2}{\bar\mu_1}(\sin\bar\mu_2c_2+
\sin\bar\mu_3c_3)+\ell_o p_2p_3\right)\cos\bar\mu_1c_1; \\
\dot{c_1} &= -\f{1}{\gamma}\Bigg[\f{p_2p_3}{\Delta\lp^2}\bigg(
\sin\bar\mu_1c_1\sin\bar\mu_2c_2+\sin\bar\mu_1c_1\sin\bar\mu_3c_3+\sin\bar\mu_2
c_2\sin\bar\mu_3c_3 \nonumber \\ & \qquad +\f{\bar\mu_1c_1}{2}\cos\bar\mu_1c_1(
\sin\bar\mu_2c_2+\sin\bar\mu_3c_3)-\f{\bar\mu_2c_2}{2}\cos\bar\mu_2c_2(\sin\bar
\mu_1c_1+\sin\bar\mu_3c_3) \nonumber \\ & \qquad -\f{\bar\mu_3c_3}{2}\cos\bar
\mu_3c_3(\sin\bar\mu_1c_1+\sin\bar\mu_2c_2)\bigg)+\ell_o\bigg(\f{3}{2\bar\mu_1}\bigg[
\f{p_1p_2}{p_3}\sin\bar\mu_3c_3
+\f{p_1p_3}{p_2}\sin\bar\mu_2c_2 \nonumber \\ & \qquad
-\f{p_2p_3}{3p_1}\sin\bar\mu_1c_1\bigg]
+\f{1}{2}\f{p_2p_3}{p_1}c_1\cos\bar\mu_1c_1
-\f{1}{2}p_2c_3\cos\bar\mu_3c_3
-\f{1}{2}p_3c_2\cos\bar\mu_2c_2\bigg) \nonumber \\ & \qquad
+\f{\ell_o^2}{4}(1+\gamma^2)\bigg(4p_1-2p_1\left(\f{p_2^2}{p_3^2}
+\f{p_3^2}{p_2^2}\right)+2\f{p_2^2p_3^2}{p_1^3}
\bigg)\Bigg]. \end{align}
The equations for $\dot{p_2}, \dot{p_3}, \dot{c_2}$ and $\dot{c_3}$ are
the same modulo the appropriate permutations.  Note that it is easy to
extend this for other matter fields and also to the vacuum case simply
by appropriately modifying the matter part of the effective Hamiltonian
constraint.

In the embedding approach these effective equations provide quantum
geometry corrections to the classical equations of motion
Eqs.~(\ref{ceom1})~and~(\ref{ceom2}) due to the area gap.  However,
careful numerical work comparing the full quantum dynamics to the
effective dynamics is necessary to determine whether the effective
equations are accurate beyond first order in $\hbar$.

Now, it is well known that classical Bianchi IX space-times with a
massless scalar field as a matter source behave in an asymptotically
velocity term dominated (AVTD) manner%
\footnote{This is true so long as the constant of motion $p_T^2$
is large enough so that the three scale factors are all decreasing
as the singularity is approached.},
that is to say that the potential term is negligible (see \cite{bb}
and references therein).  For certain regions of phase space, this
will occur \emph{before} quantum gravity effects become important
and we will assume that in this case only quantum gravity
corrections to the velocity terms are relevant.

It then follows that this behaviour is identical to that of the
Bianchi I model and therefore the effective Friedmann equation for
the Planck regime to first order in $\hbar$ is given by \cite{cv}
\be H^2 = \label{H2quant}\f{8\pi G}{3}\rho \left(1 - \f{\rho}{\rho_c}
\right) + \f{\Sigma^2}{6} - \f{\Sigma^2\rho}{2\rho_c} - \f{(\Sigma^2)^2}
{32\pi G \rho_c} + O(\lp^4), \ee
where $\rho_c = 3/8\pi\gamma^2\Delta G\lp^2 \approx 0.41 \rho_{\rm Pl}$
(recall that $\Delta = 4\sqrt{3}\pi\gamma$ and $\gamma \approx 0.2375$
due to black hole entropy calculations \cite{bhentropy}).  The expression
for $\Sigma^2$ is given by
\be \Sigma^2 = \f{1}{3\gamma^2 p^3}\Big[(p_1c_1 - p_2c_2)^2
+ (p_2c_2 - p_3c_3)^2 + (p_3c_3 - p_1c_1)^2 \Big], \ee
and one can show that $p^3\Sigma^2$ is a constant in the AVTD limit \cite{cv}.

It is clear that there is a bounce ($H^2 = 0$) when the matter energy density
reaches
\be \rho_{\rm bounce} = \f{1}{2}\left[\rho_c - \f{3\Sigma^2}{16\pi G}
+ \sqrt{\left(\rho_c - \f{3\Sigma^2}{16\pi G}\right) \left(\rho_c
- \f{\Sigma^2}{16\pi G}\right)}\right], \ee
at which point the energy density and curvature will both decrease and leave
the Planck regime and the classical dynamics will once again become a good
approximation.  It follows that the matter energy density is always bounded
above by the critical energy density $\rho_c = 0.41 \rho_{\rm Pl}$.  This is
only an upper bound as the matter density at the bounce depends quite strongly
on $\Sigma^2$ which is a measure of the strength of the gravitational waves:
the stronger the gravitational waves are, the lower $\rho_{\rm bounce}$ will be.

The scenario described above relies on the AVTD behaviour of the Bianchi IX
cosmology with a massless scalar field occuring before quantum gravity effects
become important.  In this case, the true Friedmann equation can then be well
approximated by Eq.~(\ref{H2class}) in the classical regime and by
Eq.~(\ref{H2quant}) in the AVTD limit.  However, this scenario \emph{will not}
be valid for all regions of phase space, in particular where the scalar field
momentum $p_T$ is small enough for the chaotic Mixmaster behaviour to appear.

It has been suggested that, by bounding the strength of the potential terms
due to inverse triad effects, quantum gravity effects could play an important
role in Bianchi IX dynamics and that the chaotic Mixmaster behaviour would be
avoided as a result of this for all types of matter fields \cite{bdh}.  In
the effective equations presented above, we have ignored the effect of inverse
volume corrections (which for the inverse volume operator used in this paper
are only important for $v<4$) and have only considered the effect of holonomy
corrections.  If the chaotic behaviour is to be generically avoided in this
effective theory, it will be because the repulsive quantum gravity effects
will ensure that the Bianchi IX space-time will not remain in high curvature
regions for long enough for there to occur a sufficient number of Mixmaster
bounces for chaos to appear.

For now, this remains a conjecture and one would have to study the Bianchi IX
effective equations of motion more carefully, using both analytic and numerical
methods, in order to determine whether the bounce is generic and also to see if
chaotic behaviour is avoided or not in the effective theory for small $p_T^2$.

\section{Discussion}
\label{s5}

In this paper we have studied the improved LQC dynamics for Bianchi IX
cosmologies where the matter content is a massless scalar field which
is used as a relational time parameter.  We have shown that the singularities
in the classical theory are resolved by quantum gravity effects in the
usual manner in LQC as the singular states decouple from the regular ones
under the relational dynamics given by the Hamiltonian constraint operator.

It is important to point out that all of the tools necessary for the
task of deriving the LQC dynamics for Bianchi IX models were already
available.  First, the form of $\bar\mu_i$ was introduced in the
study of Bianchi I models \cite{awe2}, as were the variables $\la_i$
which greatly simplify the form of the action of the Hamiltonian
constraint operator.  The other two necessary ingredients to the
results for this work are the connection operator and the inverse
triad operators, both of which were introduced for the study of
Bianchi II models in \cite{awe3}.  In addition, even the factor-ordering
choices necessary in the Hamiltonian constraint operator had been made
in \cite{awe2, awe3, gowdy}.  Because of this, it is reasonable to
expect that no additional machinery should be necessary in order to
complete the study of the loop quantum cosmology of the other Bianchi
models of type A.

Finally, in addition to obtaining a well-defined LQC Hamiltonian
constraint operator for Bianchi IX space-times and studying some of
its properties, we also derived some effective equations which provide
modifications to the classical equations of motion due to the area
gap which is a manifestation of quantum geometry in LQG.  Although
all of the results presented in this paper were derived for the
particular case of a massless scalar field as the matter field, it
will be easy to extend the results presented here for other types
of matter fields (as well as the vacuum case) for both the quantum
and effective theories.

Of course, it is not enough to know the form of the equations of
motion given in Eq.~(\ref{qHamfin}) in order to understand the full
dynamics of the loop quantum cosmology of Bianchi IX models.
Numerical studies will be particularly useful and help us understand
how the quantum state of a Bianchi IX cosmology evolves with time.  Most
interesting would be a study of states which are sharply peaked
around a semi-classical state at late times and to then evolve them
back in time to see what happens as the curvature increases.
Based upon previous experience with isotropic models, one might
expect to see one or several bounces as the curvature reaches
the Planck scale but careful numerical studies are needed to check
this.

If the BKL conjecture is correct a good understanding of the
quantum dynamics of Bianchi IX cosmologies will lead to a better
understanding of the behaviour of generic space-times as their
curvature reaches the Planck scale.  If Bianchi IX models are sufficiently
rich in order to understand the approach to such regions, it would
appear that no singularities would form since an initially nonsingular
Bianchi IX wave function must remain nonsingular as shown in
Sec.~\ref{s3}.  It is therefore possible that a careful study of
the BKL conjecture at the level of the quantum dynamics could provide
a no-singularity theorem, a first step in this direction is provided
by \cite{ahs}.

A simpler avenue to study quantum gravity effects in Bianchi IX models
would be to study the effective dynamics presented in Sec.~\ref{s4}.
In isotropic models it turns out that the effective dynamics are
surprisingly accurate even in the deep Planck regime: the effective
equations accurately predict the quantum trajectory throughout the
quantum bounce for sharply peaked wave functions.  Because of this,
it would be interesting to study the dynamics given by the effective
equations for Bianchi IX space-times.  However, it is essential to
see  where the effective equations break down, if they do at all.
This can be done by including higher order corrections to the
effective equations via the moment expansion method and also by
comparing the predictions of the effective equations to full
numerical solutions of the Hamiltonian constraint operator.

An analysis of the effective equations of motion in the case where the
asymptotically velocity term dominated behaviour begins before quantum
gravity effects become important shows that there is a bounce when the
curvature reaches the Planck scale and that the matter energy density
is bounded above by the critical energy density $\rho_c \approx 0.41
\rho_{\rm Pl}$.  This result relies on the AVTD behaviour and is
\emph{not} generic.  Therefore, one must also examine other areas in
the phase space in order to fully understand the predictions of the
effective theory, particularly near Planck scales.  As the Mixmaster
behaviour appears for small $p_T^2$ during the approach to the
singularity in the classical theory, the effective equations can
provide a better understanding of how quantum gravity effects may modify
the Mixmaster behaviour as well.  In particular it is possible that, as
for simpler isotropic models and in AVTD case here, these quantum gravity
effects will be repulsive and cause a quantum bounce.  This would limit
the amount of time that the Mixmaster behaviour occurs and the chaos
which arises in the classical theory might be avoided due to the short
time span of the Mixmaster dynamics.  However, this remains a conjecture
and much more work, both analytic and numerical, is needed in order to
resolve this question.

Finally, it has been pointed out that the dipole cosmology model can
be used in order to study the Bianchi IX model \cite{bmr}.  Although
that paper studies the Euclidean theory, it would nonetheless be
interesting to compare the model presented in \cite{bmr} with the one
developed in this paper.  In particular, \cite{bmr} suggests two possible
approaches in order to obtain the Hamiltonian constraint operator for
their model.  Comparing the quantum dynamics resulting from these two
possibilities to those derived in this paper could help determine which
of the two approaches is the correct one and hence give some insight
into the dipole cosmology models and also spin foam models in general.
It is also possible to further probe the relation between the canonical
and the covariant approaches to LQG via LQC by extending the Feynman
path integral construction given in \cite{ach} for the flat FRW model
to the Bianchi IX model; this extension would be nontrivial due to the
additional degrees of freedom present in Bianchi IX space-times, but it
could also improve our understanding of the connection between the
canonical and covariant approaches to LQG as well as the relation
between full LQG and the symmetry-reduced models of LQC.

\section*{Acknowledgements:}

The author would like to thank Abhay Ashtekar and Martin Bojowald
for helpful discussions. This research was supported in part by
NSF grant PHY0854743, the George A. and Margaret M. Downsbrough
Endowment, the Eberly research funds of Penn State, Le Fonds
qu\'eb\'ecois de la recherche sur la nature et les technologies and
the Edward A. and Rosemary A. Mebus funds.

\end{document}